# Galactic Spiral Structure

CHARLES FRANCIS, ERIK ANDERSON

We describe the structure and composition of six major stellar streams in a population of 20574 local stars in the New Hipparcos Reduction with known radial velocities. We find that, once fast moving stars are excluded, almost all stars belong to one of these streams. The results of our investigation have lead us to re-examine the hydrogen maps of the Milky Way, from which we identify the possibility of a symmetric two-armed spiral with half the conventionally accepted pitch angle. We describe a model of spiral arm motions which matches the observed velocities and composition of the six major streams, as well as the observed velocities of the Hyades and Praesepe clusters at the extreme of the Hyades stream. We model stellar orbits as perturbed ellipses aligned at a focus in coordinates rotating at the rate of precession of apocentre. Stars join a spiral arm just before apocentre, follow the arm for more than half an orbit, and leave the arm soon after pericentre. Spiral pattern speed equals the mean rate of precession of apocentre. Spiral arms are shown to be stable configurations of stellar orbits, up to the formation of a bar and/or ring. Pitch angle is directly related to the distribution of orbital eccentricities in a given spiral galaxy. We show how spiral galaxies can evolve to form bars and rings. We show that orbits of gas clouds are stable only in bisymmetric spirals. We conclude that spiral galaxies evolve toward grand design two-armed spirals. We infer from the velocity distributions that the Milky Way evolved into this form about 9Gyrs ago.

## 1. Background

Stellar orbits are not elliptical because the gravitating mass of the galaxy is distributed in the disc and the halo (Binney & Tremaine, 1987, chapter 3). In addition, orbits oscillate in the direction perpendicular to the disc. Orbits are expected to precess from both these causes, generating a rosette. It is usually assumed that, in time, an equilibrium state will be attained in which the distribution is well-mixed.

Spiral structure is usually explained using the density wave hypothesis of Lin et al. (1969), according to which stars move through the arms, which consist of dense regions analogous to regions of heavy traffic on a motorway. A simple analogy with patches of heavy traffic fails because a wave effect would require that stars slow down when they approach a dense region, but the gravity of the dense region would cause them to speed up. Following Kalnajs (1973), density wave theory is usually explained by means of a diagram (such as figure 1) constructed by enlarging and rotating ovals. The orbit is an epicyclic approximation in coordinates rotating at a rate $\Omega - K/2$, where $\Omega$ and $K$ are the frequencies of circular motion and the radial oscillation (see appendix A).

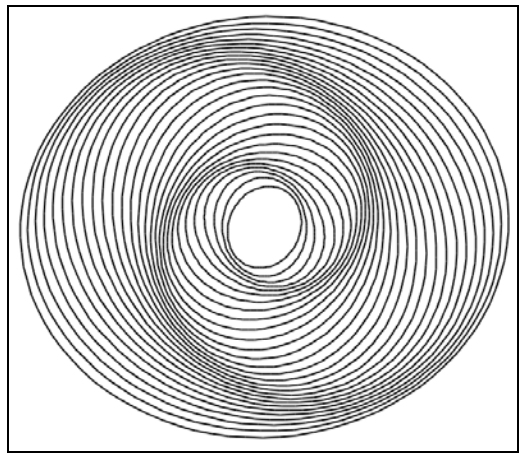

**Figure 1:** Enlarging and rotating ovals aligned at the centre generates a two-armed spiral.

Density wave theory could apply to gas in laminar flow, but not to observed turbulent gas motions, or to mature stars for which the increase in density in figure 1 represents only a small proportion of the orbit. If, as one expects, $K$ is similar in magnitude to $\Omega$ (see section 7), then coordinates rotate at near half orbital velocity and figure 1 shows a single spiral twice, not a bisymmetric spiral. Orbits which appear separate in these coordinates, and which are treated as a laminar flow in density wave theory, actually cross in physical space. We show in appendix A that an entire class of rotating coordinates in which orbits are closed has been overlooked in Lindblad's epicycle theory, including the most natural solution. Other models of spiral arms include the stochastic self-propagating star formation hypothesis proposed by Mueller & Arnett (1976), which suggests a mechanism by which young stars may form transitory spiral segments, but does not apply to enduring grand design two-armed spirals.

The established models are not borne out by the analysis of kinematic data in the solar neighbourhood. After the removal of fast moving stars, whose orbits are highly eccentric and/or significantly inclined to the disc, and which principally belong to the halo or the thick disc, far from being well-mixed or obeying a laminar flow, the remaining population of thin disc stars divides into six major streams with distinct motions, and containing stars of all ages. We will describe an alternative mechanism, which does not depend on epicycles, and which also results in spiral structure. We will show that this structure is dynamically stable, and that the observed stream motions are precisely those which the structure predicts.

## 2. Stellar Streams

Stellar streams consist of large populations of stars with similar motions. The existence of moving groups was first established from astronomical investigations dating as far back as 1869 (Eggen, 1958). They were thought to consist of previously clustered coeval stars that have been gradually dispersed by the dynamic processes of tidal forces, differential galactic rotation, and encounters with other stars. Increasingly comprehensive star catalogues published in the 1950's opened the way for more thorough analyses. Beginning in 1958, Eggen produced a series of seminal studies of stellar streams using RA:DE proper motion ratios in conjunction with radial velocities. Eggen's investigations showed significantly increased membership counts and spatial extents of stellar streams, leading him to hypothesize a more protracted process of dissolution for star clusters. In Eggen's scenario, as star clusters dissolve during their journeys around the Galaxy, they are stretched into tube-like formations, which were subsequently called superclusters.

The investigation of stellar streams received a major boost with the arrival of the precision astrometry afforded by the Hipparcos mission. Dehnen (1998), using transverse velocities derived from Hipparcos, produced maps of the local stellar velocity distribution showing that streams contain a significant proportion of late type stars. A wide range of stellar ages was identified within superclusters, challenging Eggen's hypothesis of common origin (e.g., Chereul et al., 1998, 1999). The search for other types of dynamical mechanisms to account for streams has been ongoing. Candidates include migrations of resonant islands (Sridhar & Touma, 1996; Dehnen, 1998) and transient spiral waves (De Simone et al., 2004; Famaey et al., 2005) in which streams originate from perturbations in the gravitational potential associated with spiral structure.

### 3. The Local Standard Of Rest

The local standard of rest (LSR) is defined to mean the velocity of a circular orbit at the Solar radius from the Galactic centre. The definition idealizes an axisymmetric galaxy in equilibrium, ignoring features like the bar, spiral arms, and perturbations due to satellites. An accurate estimate of the LSR is required to determine parameters like the enclosed mass at the solar radius and the eccentricity distribution which is of importance in understanding galactic structure and evolution.

As is customary in kinematic analyses of the stellar population, we denote velocity in the direction of the galactic centre by $U$, in the direction of rotation by $V$, and perpendicular to the galactic plane by $W$. The solar motion relative to the LSR is $(U_0, V_0, W_0)$. The usual way to calculate the LSR is to calculate the mean velocity of a stellar population, and to correct $V_0$ for asymmetric drift. The method assumes a well-mixed distribution. However, as is seen in section 4, the observed kinematic distribution is highly structured, and divides into six populations each with distinct motion and stellar composition. Ignoring the possibility of perturbations to the galactic plane, motions of thin disk stars in the $W$-direction may be treated as a low amplitude oscillation due to the gravity of the disc, and as independent of orbital motion in the $U$-$V$ plane. It is thus not unreasonable to calculate $W_0$ as the mean motion of a population. However, in the absence of knowledge of the causes for streams, there is no way to relate the statistical properties of their motion to $U_0$ and $V_0$.

Francis and Anderson (2009) studied a population of 20 574 Hipparcos stars with complete kinematic data, described in appendix B. We observed a deep minimum in the velocity distribution at a particular value of $(U, V)$ and argued that such a minimum might be expected at the LSR as a consequence of disc heating. Heating is the process by which scattering events cause the random velocities of stars to increase with age (e.g., Jenkins, 1992). In thermal equilibrium in a well-mixed population, one would expect that the modal magnitude of random peculiar velocity denotes disc temperature. Circular motion represents an absolute zero temperature and can be expected to be rare for mature orbits. As a result, the distribution in velocity space can be expected to have a minimum at circular motion. In this paper we will show that the true cause of the minimum is the perturbation of orbits due to spiral structure. We will use the value of the solar motion found from the minimum in the velocity distribution, $(U_0, V_0, W_0) = (7.5 \pm 1.0, 13.5 \pm 0.3, 6.8 \pm 0.1)\,\mathrm{km\,s^{-1}}$. We will use an adopted Solar transverse orbital velocity of 225 kms$^{-1}$ and a distance to the Galactic centre of 7.4 kpc, consistent with recent determinations (Reid, 1993; Nishiyama et al., 2006; Bica et al., 2006; Eisenhauer et al., 2005; Layden et al., 1996) and the proper motion of Sgr A* determined by Reid and Brunthaller (2004) on the assumption that Sgr A* is stationary at the Galactic barycentre.

### 4. Stream properties

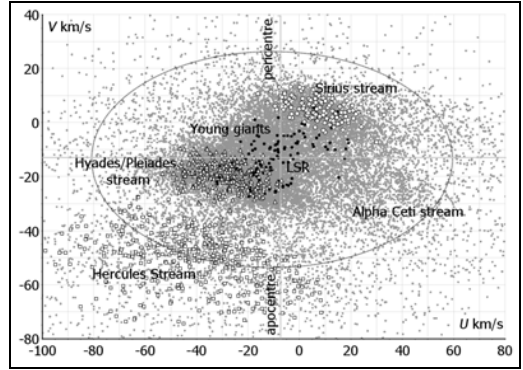

Famaey et al. (2005) described six kinematic groups: three streams, Hyades/Pleiades, Sirius and Hercules, a group of young giants, high velocity stars and a smooth background distribution (figure 2). We smoothed the velocity distribution by replacing each discrete point with a two-dimensional Gaussian function and finding the sum. The choice of smoothing parameter depends on the density of stars in the plot, and the required visual balance between overall structure and detail. Too large a smoothing parame-ter obscures structure, while two small a value confuses random fluctuations with structure. A standard deviation of $0.75\,\mathrm{km\,s^{-1}}$ gives a clear contour plot (figure 3) and shows the major features of the distribution.

**Figure 2:** *U-V* plot showing groups identified by Famaey et al. (2005). These represent only a small proportion of the true membership of the streams. The calculated position of the LSR is shown for clarity.

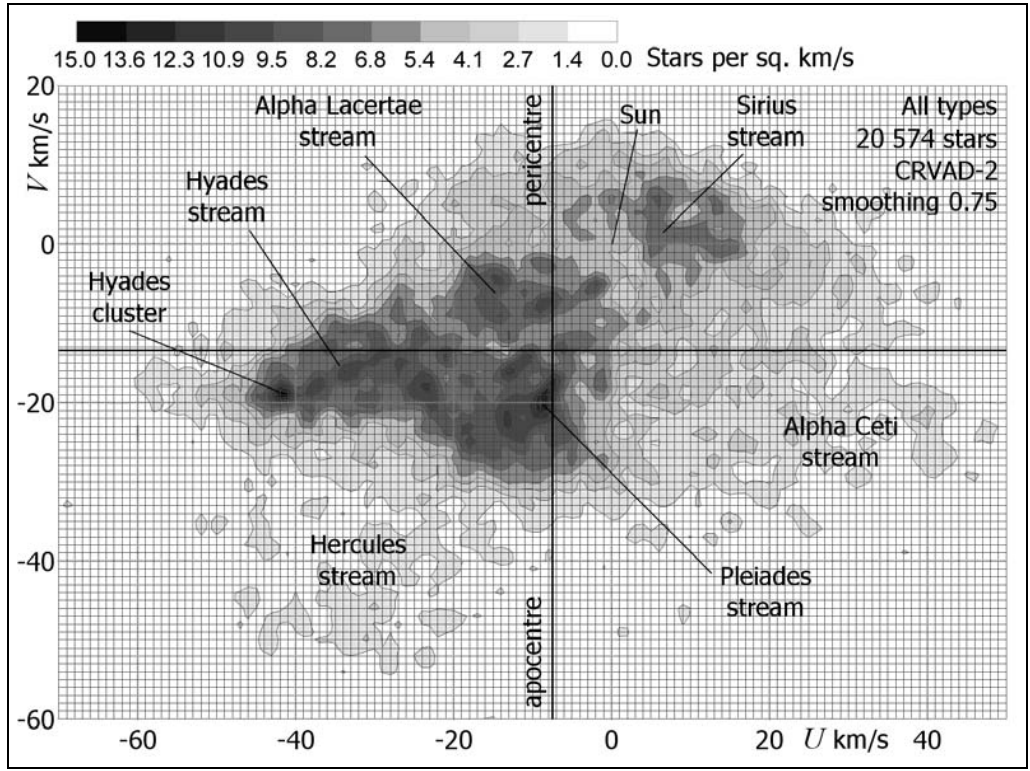

**Figure 3:** The distribution of *U*- and *V*-velocities using Gaussian smoothing with a standard deviation of 0.75 kms$^{-1}$, showing the Hyades, Pleiades, Sirius, Hercules, Alpha Lacertae and Alpha Ceti streams.

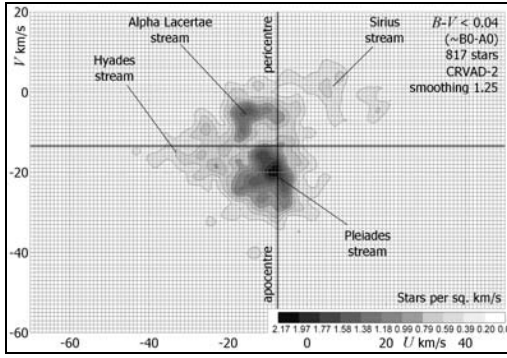
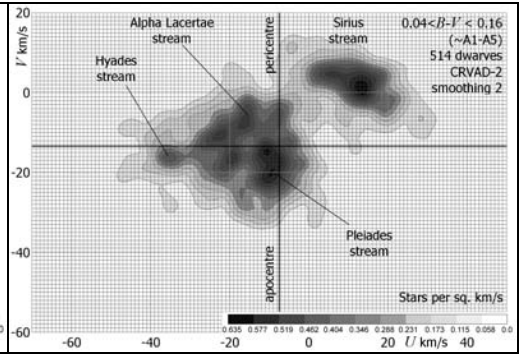

**Figure 4:** *U-V* distribution: 817 dwarfs, $B-V < 0.04$ (~B-A0), smoothing $\sigma = 1.25$.

**Figure 5:** *U-V* distribution: 514 dwarfs, $0.04 \leq B-V < 0.16$ (~A1-A5), smoothing $\sigma = 2$.

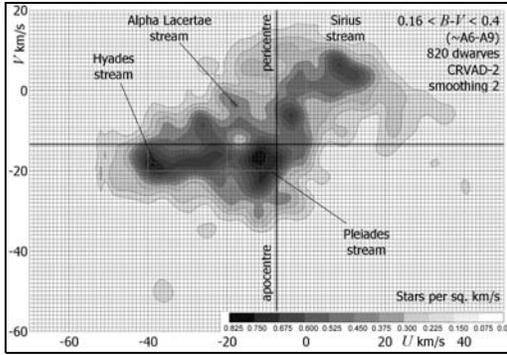
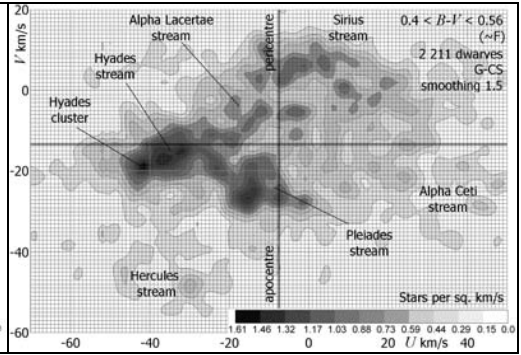

**Figure 6:** *U-V* distribution: 820 dwarfs, $0.16 \leq B-V < 0.4$ (~A6-A9), smoothing $\sigma = 2$.

**Figure 7:** *U-V* distribution: 2211 G-CS dwarfs, $0.4 \leq B-V < 0.56$ (~F), smoothing $\sigma = 1.5$.

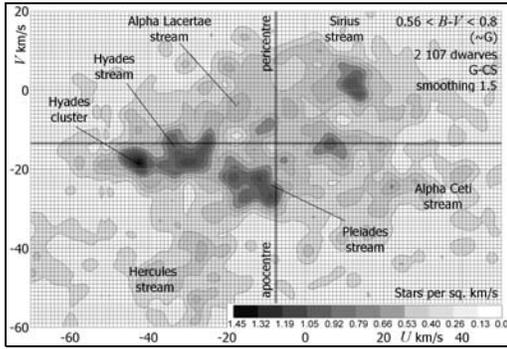
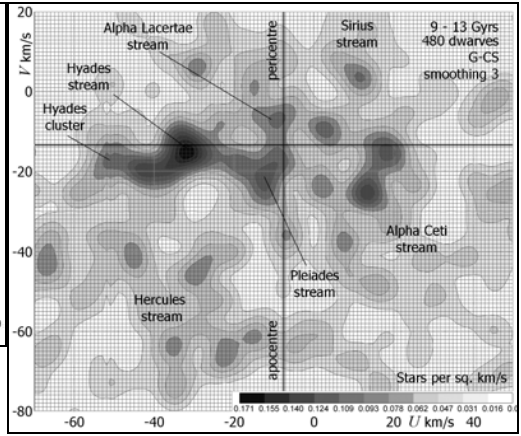

**Figure 8:** *U-V* distribution: 2107 G-CS dwarfs, $0.56 \leq B-V < 0.8$ (~G), smoothing $\sigma = 1.5$.

**Figure 9:** *U-V* distribution: 480 G-CS dwarfs with ages 9-13 Gyrs, smoothing $\sigma = 3$.

Streams are seen as dense regions in the velocity plots (figure 3 to figure 9). It is not possible to give precise criteria for stream membership from purely statistical data, as there is some overlap. Here we seek only a broad description. We distinguish the Hyades and Pleiades streams, since the velocity distributions shows separate peaks, and, as we will see, these streams contain different distributions of stellar types and ages. There is a large and well dispersed stream centred at $(U, V) = (25, -23)\,\mathrm{km\,s^{-1}}$, noted by Dehnen (1998). Our estimate of its position is in good agreement with Chakrabarty (2007) who identified a clump in the velocity distributions at $(U, V) \approx (20, -20)\,\mathrm{km\,s^{-1}}$. We have called it the Alpha Ceti stream, after the brightest star we identified with this motion.

Our analysis shows that the Pleiades stream consists largely of new-born stars, originating in our own spiral arm and with low eccentricities and typical orbits near to apocentre, and also contains mature orbits with slightly greater eccentricity. We distinguish it from a stream with orbits close to pericentre which contains young as well as old stars. We have called this the Alpha Lacertae stream. It appears that Famaey's young giants belong to the Alpha Lacertae stream. Famaey found a total stream membership of over 25%, but the velocity distribution is highly structured by colour. When this is taken into consideration one sees that streams represent the bulk of the population.

The bluest stars, with $B - V < 0.04$ (~B-A0) reflect recent star formation in the Pleiades and Alpha Lacertae streams (figure 4). The overlap between these streams and the Hyades stream makes it difficult to ascertain the earliest Hyades stars. There are Hyades candidates of type B2, having a maximum age of about 20 Myrs, stronger candidates at type B4, a maximum age of about 60 Myrs, and clear signs of a Hyades population at type B9, an age of about 400 Myrs. Apart from a single star of type B7, the earliest clear indication of the Sirius stream is for stars of type B8, an age about 300 Myrs. The Hercules and Alpha Ceti streams also contain members as early as B8, but become well populated at type F0, corresponding to an age of about 2½ Gyrs.

For $0.04 \leq B - V < 0.16$ (~A1-A5), the velocity distribution is concentrated in the Pleiades and Sirius streams (figure 5). The Hyades stream becomes more prominent than the Sirius stream for $0.16 \leq B - V < 0.4$ (~A6-A9) (figure 6), and dominates the velocity distributions (by density, not by total population) for dwarves with $0.4 \leq B - V < 0.56$ (~F) (figure 7) and $0.56 \leq B - V < 0.8$ (~G).

A clear indication of the stability of stream motions is given by the velocity distribution for old stars, using isochrone ages given by G-CS II (figure 9). There are known problems with isochrone aging for very young stars; we found that a number of stars with young kinematics had been assigned ages greater than 13 Gyrs. In other respects, G-CS II isochrone ages appear to be at least broadly reasonable, in accordance with position on the H-R diagram. For stars aged 9-13 Gyrs, there is little indication of Sirius or Pleiades streams. The Hyades stream shows a sharp peak. The Hercules and Alpha Ceti streams, which are diffuse but contain more stars, are also prominent.

## 5. The Eccentricity Distribution

For an elliptical orbit the eccentricity vector is defined as the vector pointing toward pericentre and with magnitude equal to the orbit's scalar eccentricity. It is given by

$$\boldsymbol{e} = \frac{|v|^2 \boldsymbol{r}}{\mu} - \frac{(\boldsymbol{r} \cdot \boldsymbol{v})\boldsymbol{v}}{\mu} - \frac{\boldsymbol{r}}{|\boldsymbol{r}|},$$

where $\boldsymbol{v}$ is the velocity vector, $\boldsymbol{r}$ is the radial vector, and $\mu = GM$ is the standard gravitational parameter for an orbit about a mass $M$ (e.g. Arnold, 1989; Goldstein, 1980). For a Keplerian orbit the eccentricity vector is a constant of the motion. Stellar orbits are not strictly elliptical, but the orbit will approximate an ellipse at each part of its motion, and the eccentricity vector remains a useful measure (the Laplace-Runge-Lenz vector, which is the same up to a multiplicative factor, is also used to describe perturbations to elliptical orbits). We smoothed the eccentricity distribution by replacing each discrete point with a two dimensional Gaussian function and finding the sum. Standard deviation, $\sigma$, is used as a smoothing parameter. A standard deviation of 0.005 gave a clear contour plot (figure 10). In a well-mixed population, eccentricity vectors will be spread smoothly in all directions, with an overdensity at apocentre and underdensity at pericentre, because of the increased orbital velocity at pericentre and because stars at apocentre come from a denser population nearer the galactic centre. This is not seen in the plot. In practice the distribution is concentrated at particular values, corresponding to stream motions.

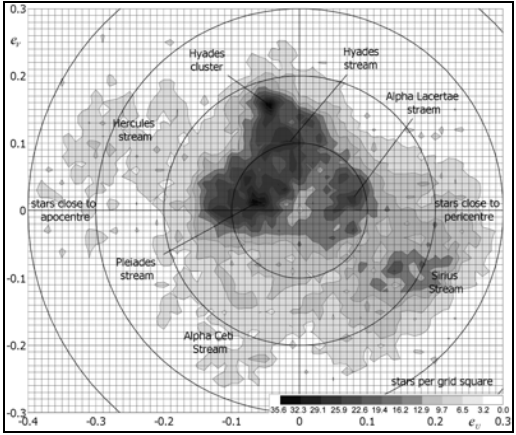

**Figure 10:** Contour of the density of the eccentricity distribution, based on the value of the LSR found in Francis & Anderson, 2009. $e_U$ and $e_V$ are the components of the eccentricity vector toward the galactic centre and in the direction of rotation. The Hercules stream has eccentricities up to ~0.3 and orbits approaching apocentre. The Sirius and Alpha Ceti streams have eccentricities ~0.1-0.25 approaching pericentre. The Hyades stream has eccentricities below ~0.2 approaching apocentre. The Pleiades stream has typical eccentricities about ~0.06 close to apocentre.

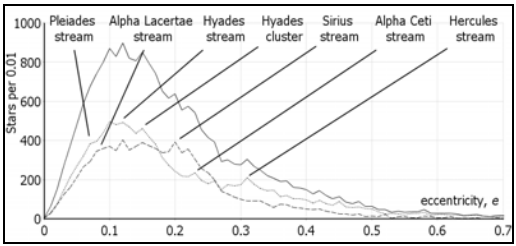

**Figure 11:** Eccentricity distribution (based on the LSR found in Francis & Anderson, 2009) for the entire population, for stars closer to apocentre (dotted) and stars closer to pericentre (dashed), as defined by position with respect to the semi-latus rectum.

## 6. A Model of Spiral Structure

In polar coordinates $(r,\theta)$ an equiangular spiral is given, for positive real $a$ and $b$, by

$$r = ae^{b\theta}.$$

The constant pitch angle of the spiral is $\phi = \operatorname{atan} b^{-1}$. We demonstrate here that an equiangular spiral structure can be constructed from elliptical orbits by enlarging an ellipse by a constant factor, $k$, centred at the focus and rotating it by a constant angle, $\tau$, with each enlargement (figure 12). Orbits of difference sizes align in a spiral pattern, leading to an overdensity of stars which creates the spiral arms. The pitch angle of the spiral depends only on $k$ and $\tau$, not the eccentricity of the ellipse. For a given pitch angle, ellipses with a range of eccentricities can be fitted to the spiral, depending on how narrow one wants to make the spiral structure and what proportion of the circumference of the ellipse one wants to lie within it. In general terms, ellipses with greater eccentricity fit with spirals with greater pitch angles.

In a practical model for galactic spiral structure, stellar orbits are approximately elliptical and are gravitationally aligned to a spiral arm. Unaligned orbits lying between the arms will be drawn to one arm or the other, and orbits will precess due to the distributed matter distribution of the galaxy, such that they become aligned. Once alignment of the orbit with the spiral arm is achieved, it will be maintained by perturbations to the orbit due to the gravity of the arm. A star close to apocentre will approach the inside of the arm, on account of the pitch angle. If it has greater eccentricity than that of stars in the arm, the gravity of the arm will draw it closer, causing a reduction in eccentricity. If it has lower eccentricity than the arm stars, it will pass through the arm. Because of the curve of its orbit, it will spend more time in the gravitational field on the outside of the arm, and will be drawn back towards the arm, with a net increase in eccentricity.

The mechanism binding stars to the arm is explained in more detail in appendix E. It reinforces the spiral, showing that spiral arms are stable dynamical structures (they may eventually be destroyed by the growth of a bar and/or a ring). Orbital precession will mean that the spiral pattern rotates, but the winding problem is resolved because, for a wide range of orbits, orbital alignment is determined by the gravitational field of the arm (the classic winding problem is inapplicable because spiral structure does not depend directly on orbital velocity at different radii). The evolution of bisymmetric spirals from flocculent and multi-armed spirals is explained in section 12. The model predicts trailing spirals, not leading spirals, because orbital velocity and gravitational field strength due to a centrally concentrated mass distribution are lower near apocentre, so that the gravitational field of the arm is of greater influence in perturbing the orbit near apocentre. Thus the alignment of orbits to the spiral arm proceeds from the outside toward the inside, not the other way about. This agrees with the long established observational result (de Vaucouleurs, 1958). The few exceptional candidates for leading spirals are thought to be induced by special mechanisms of tidal interactions with companion galaxies. (Väisänen et al., 2008).

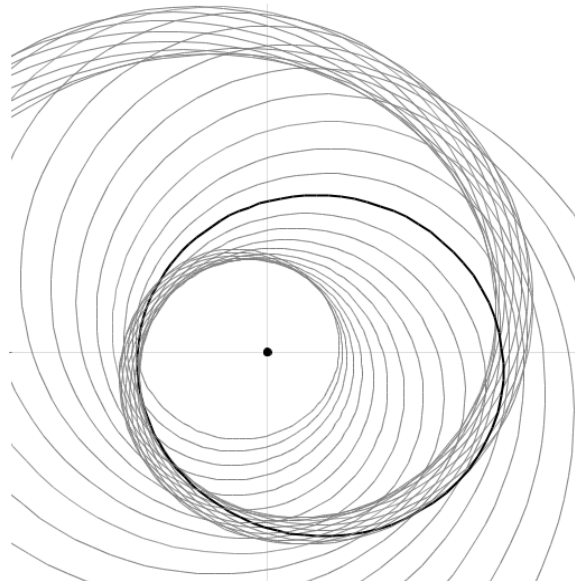

**Figure 12:** An equiangular spiral with a pitch angle of 11°, constructed by repeatedly enlarging an ellipse with eccentricity 0.3 by a factor 1.05 and rotating it through 15° with each enlargement. Lower eccentricity ellipses produce a narrower structure. Ellipses with eccentricity greater than about 0.25 have more than half their circumference within the spiral region. Ellipses with eccentricity greater than about 0.35 produce probably too broad a structure to model a spiral arm with this pitch angle, but give a good fit for spirals with greater pitch angle.

An animation of a galaxy formed from aligned rosettes with similar parameters to the Milky Way is described in appendix C. The animation clearly shows stars crossing an arm at the same part of their outward motion (Hyades stream), as well as the differing velocities of stars in the arm.

## 7. Precession of Apocentre and Spiral Pattern Speed

For an orbit in the thin disc, motion perpendicular to the disc may be treated as an independent oscillation superimposed on an orbit in the plane of the disc. To a good approximation, this oscillatory motion does not cause the orbit to precess (since the oscillation is perpendicular to the centripetal force and to orbital motion). The mass distribution in the halo is generally assumed spherical. By Newton's shell theorem it can be treated as a central mass, which reduces as orbital radius reduces. The effect is to reduce the curvature of the orbit at pericentre, such that pericentre regresses during the orbit. The result is less obvious for matter in the disc, because the gravitational effect of nearby matter in a uniform ring outweighs the net effect of the further parts of the ring. The reduction in curvature of the orbit near pericentre due to lower enclosed mass is offset by the increase in density of nearby matter. We used a numerical simulation for a galaxy with a central core plus a disc with an exponentially decreasing surface density to establish that orbits also regress (figure 13).

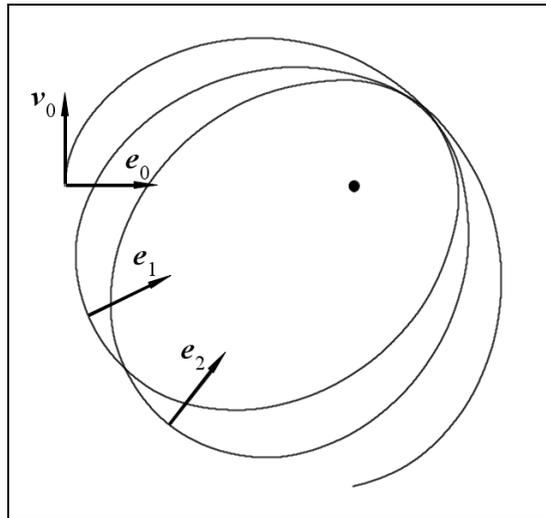

**Figure 13:** The eccentricity vector of an orbit regresses for a central core plus disc. Regression has been exaggerated by increasing the mass of the disc relative to the core (by comparison with the Milky Way). The simulation used a central mass of 35 billion solar masses, a disc density $0.3e^{-R/3}$ billions solar masses per $kpc^2$, initial radius 8kpc and initial velocity $190 \text{kms}^{-1}$.

Knowledge of the mass distribution of the Milky Way is not sufficiently precise to choose a specific model from which an exact rate of precession may be calculated, but it is not large. The distribution of dark matter in the halo and in the thin disc affects the rate of precession of stellar orbits, but it has no direct impact on spiral structure. If we assume that the rate of precession is constant, or approximately so, for orbits at different radii, and for eccentricities within the range of binding by the spiral arm, then we may choose coordinates rotating at the rate of precession of apocentre. In these coordinates orbits do not precess, and may be taken to be approximately elliptical. The spiral arm structure applies as before, and we find that, in non-rotating coordinates, spiral pattern speed is equal to the rate of precession of apocentre.

We only require that the rate of precession is an approximate constant for a stable spiral arm, because gravitational binding to the arm outweighs the consequence of small changes in rate of precession at different orbital radii and for different eccentricities. If the matter distribution is such that orbital precession is not constant for different radii, the pitch angle of the spiral will alter over time. Eccentricities will adjust to the pitch angle, and stability may be achieved at altered values of pitch angle and eccentricity (we may conjecture that the rings of Saturn are comprised of very tightly wound spirals built from orbits which owe their low eccentricity to the mass distribution). For the purpose of this paper we will use coordinates rotating at spiral pattern speed, and we will assume that orbits can be approximated by ellipses in these coordinates.

### 8. Fitting the Model to the Milky Way

It is straightforward to observe spiral structure in other galaxies, but extremely difficult to observe it within our own galaxy, as recently illustrated by observations of the Spitzer telescope showing that stellar concentrations are not found at the positions where two arms were thought to be (Benjamin, 2008). There have been two principle methods for locating spiral arms. The usual four-armed spiral is derived principally from the distribution of ionized hydrogen (Georgelin & Georgelin, 1976; Russeil, 2003), but in fact the distribution is so sparse and irregular that it is difficult to be certain that anything has really been fitted. We will see in section 9 that ionised hydrogen is not expected to give a good fit to spiral structure in this model. The neutral hydrogen distribution was famously mapped by Oort, Kerr and Westerhout (1958), and more recently by Levine, Blitz and Heiles (2006). Levine, Blitz & Heiles fit (slightly irregular) four-armed spirals, but comment that other fits are possible.

The four-armed spirals fitted by Georgelin & Georgelin, Russeil and Levine, Blitz and Heiles have a pitch angle of about 10-15°, corresponding to orbital eccentricities in the range greater than about 0.25. This is not consistent with the eccentricity distribution for the Milky Way, in which the modal value is a little above 0.1 (figure 11), suggesting a much lower pitch angle than is used in four-armed spirals. We found a good visual fits to the hydrogen maps of Oort et al. and of Levine et al. for bisymmetric spirals with a 8.2 kpc bar and pitch angles in the range $5.3 \pm 0.5°$ (figure 14), in agreement with $5.1°$ and $5.3°$ found from HII regions for the two-armed logarithmic model by Hou, Han and

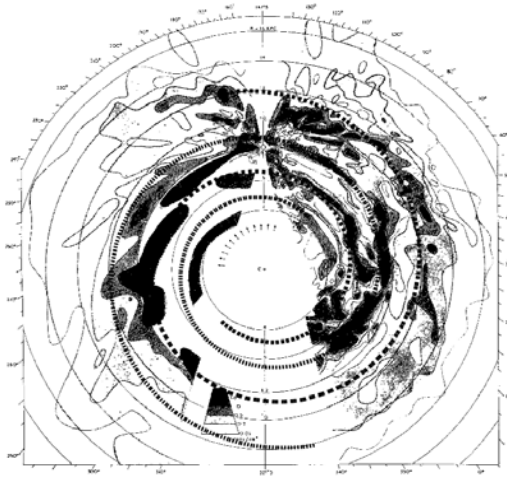

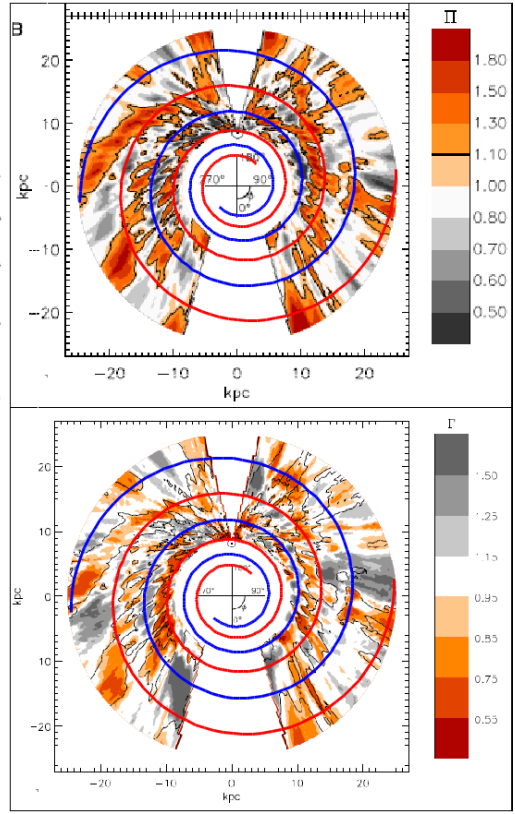

**Figure 14:** Axisymmetric logarithmic double spiral with pitch angle 5.44° and 8.2kpc bar fitted to Hydrogen maps of Oort et al and Levine et al. The spiral represents the line of the inside of the arm. The spiral structure in the Hydrogen gas distribution continues inside the radius of the bar (not plotted). The top plot from Levine et al shows density, the second shows height.

Shi (2009). There is a subjective element in the quality of such a fit, but the two-armed spirals seem to us to better follow the line of the hydrogen clouds, while the more open four-armed spirals appear to follow clouds bridging the true line of the arms. We also fitted to the map of Levine et al by maximising the mass per unit length lying on a bisymmetric spiral, finding maxima at pitch angles of 4.9° and 5.7°. As the model is predicted to give a ragged arm in gas motions, this did not appear to be better than the visual fitting method.

We constructed a symmetric two-armed spiral with pitch angle 5.44° from ellipses using an angular increment $\tau = 30°$ for each 105% enlargement (figure 19, in appendix D). For the calculated value of the LSR, current solar eccentricity is 0.138. The Sun is 16.3° before pericentre (as determined by the current eccentricity vector), at which point it should lie near the inner edge of the arm, and be heading outwards through the arm. We were not able to make a meaningful map showing the positions of stars with velocities in the arm with respect to the Sun, because the data from Hipparcos is not sufficiently comprehensive over large enough distances, and because the data for which we have radial velocities is strongly weighted to the northern hemisphere, but, in agreement with typical estimates, there is some indication in the data that we are in the arm, about 100-150pc from the inner rim, and too far to be able to detect the outer rim.

## 9. Young Stars

Star formation has been a central problem for galactic dynamics. There is not enough mass in the disc for gas clouds to collapse under gravity and form protostars. Depending on the width of the arms, orbital alignment in spiral arms results in an increase in stellar density by a factor of about 5. Gas clouds follow orbits following spiral arms according to the same laws as those governing stars (figure 15). Gas in the arm seeks to gain velocity as it approaches pericentre, and also to follow paths crossing within the arm. Thus, motions are complicated by collisions between clouds and resulting turbulence. The increase in surface density of gas is about half as much (figure 14). When clouds of atomic hydrogen with spiral arm motions meet with clouds crossing the arm it is to be expected that higher densities obtain, the height of gas is increased (figure 14), and that greater turbulence is created, with pockets of high density generating molecular gas clouds from which protostars form.

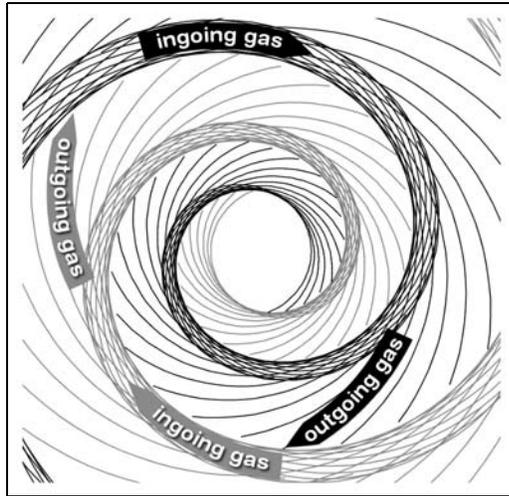

**Figure 15:** Gas clouds follow similar motion to stars, but outgoing gas is obstructed when it meets an arm.

Since outgoing gas has generally lower density than ingoing gas, outgoing motions normally terminate at the arm. In regions where outgoing gas clouds of greater than normal density meet regions where gas in the arm is less dense, the pattern of the clouds deviates from the stellar spiral, as seen in figure 14. We therefore do not expect the distribution of ionized hydrogen or of star forming regions to accurately depict spiral structure, as assumed by Georgelin & Georgelin (1976) and Russeil (2003).

We may expect that star formation typically initiates with a build-up of gas and dust on the inside of the arm, and continues through the arm. The less massive outgoing gas clouds add a component of radial velocity to those in the arm, with the result that stars form with motions found in the Pleiades stream. The modal value of eccentricity in the Pleiades stream is ~0.065, and is increasing (figure 21 in appendix E)). The first plot in figure 16 shows an orbit starting at apocentre with eccentricity 0.074. In the second, initial eccentricity is lower, 0.033. The orbits hug the outside of the arm, while eccentricity continues to increase. This is consistent with observations on many spiral galaxies showing that the brightest spots, groups of young stars, lie on the outside of the arms (e.g., M51, M74, M83, M101).

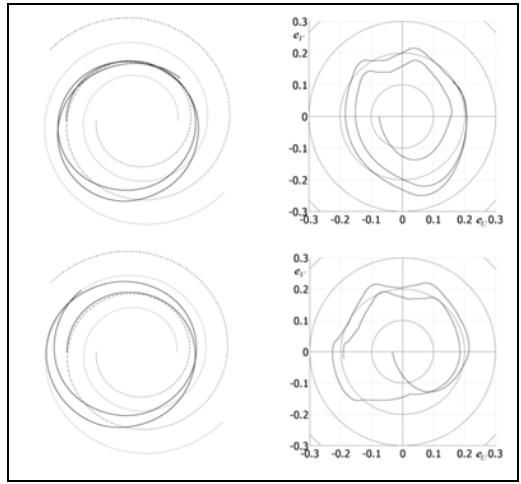

**Figure 16:** Orbits of young stars passing through apocentre with eccentricities 0.074 and 0.033, typical of the Pleiades stream

The resulting orbits achieve higher than normal eccentricity, seen in young stars in the Sirius stream and Hyades cluster. The orbits do not at first align with the arm. A typical orbit with eccentricity 0.075 at apocentre meets the arm before pericentre, with an eccentricity characteristic of the Sirius stream. Thereafter the trajectory continues to meet the spiral at pericentre for a number of orbits, while pericentre regresses so as to improve the alignment of the orbit with the arm. For eccentricity 0.034 at apocentre, the orbit does not rejoin the arm, but eccentricity continues to increase and pericentre advances, toward alignment with the other arm. In both cases it may take several more orbits before alignment with an arm is achieved.

## 10. The Hercules and Alpha Ceti streams

It is clear that the Hercules stream, and the higher eccentricities seen in the Alpha Ceti stream, do not fit with the pattern of typical spiral arm motions. Also observed on the eccentricity distribution for mature orbits (figure 23 in appendix E) are small numbers of stars with eccentricities up to about 0.3 approaching the semi-latus rectum on the inward arm of the orbit, and extending from well before pericentre to just after pericentre. These motions can also be picked out on the velocity plot (figure 22 in appendix E), and have increased prominence on the distributions for late type stars (figure 8) and old stars (figure 9), suggesting that these are also stable orbits in coordinates rotating with the arms.

It is possible to align orbits with eccentricities in the region of 0.3 with spiral arms such that the orbit follows the locus of one arm for a period after apocentre, and with the other arm during the inner part of the orbit (figure 17). Since these streams consist of predominantly older stars, we may conclude they are stable orbits formed during more turbulent motions in the early galaxy, and that young stars rarely join these motions.

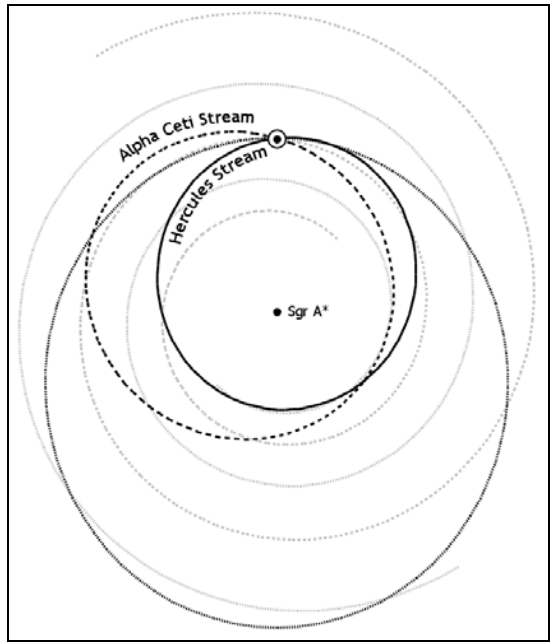

**Figure 17:** Three orbits for stars passing through the locality of the Sun with eccentricity 0.29. The continuous line shows orbits in the Hercules stream. The dashed line show orbits in the Alpha Ceti stream, and the dotted line represents orbits near pericentre in the vicinity of the Sun.

**Figure 18:** With $\mu_{arm} = 5 \times 10^8 \, m^2 s^{-2}$ the orbit continues to follow the arm after pericentre, eventually leaving with increased eccentricity; the spiral structure is broken. The initial symmetrical pattern of inter-linked orbits may not be stable. The structure may collapse to form a bar, an inner ring or both.

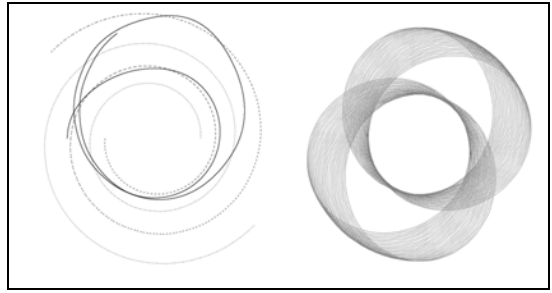

## 11. Formation of Bars and Rings

When the mass of the arm is increased in the numerical simulation, the orbit continues to follow the arm after pericentre, eventually leaving with an increased eccentricity. The density of the arm increases with the density of the disc towards the galactic centre, and, depending on the mass distribution of the galaxy, spiral structure breaks down. If, as may be expected, orbits depart from the arm at a similar point, then a symmetrical interlinked pattern of opposing orbits is formed (figure 18; e.g., UGC12646, ESO 325-28, NGC 2665, NGC 619). It seems unlikely that a structure of interlinked rings is stable. If eccentricities continue to increase and the structure collapses, a bar will be formed.

According to this model spiral arms do not start at the bar, but extend inwards beyond its ends. This is observed in a number of barred spirals (e.g., M92, M109, NGC 1300) and in the Oort hydrogen map (figure 14), in which the spiral structure of both arms continues a quarter turn beyond the bar. Once a bar is formed, it perturbs nearby orbits such that the arms are destroyed shortly after stars pass the ends of the bar. This is likely to cause the bar to grow. As with spiral arms, bar pattern speed will depend on the eccentricity of orbits in the bar. Since eccentricities in the bar are very high we may expect that bar pattern speed does not match spiral pattern speed.

In time, orbits approaching the bar will also be perturbed and will not rejoin the spiral arm. We have not performed a dynamical analysis, but conjecture that these perturbations lead to the inner ring observed in galaxies like M95, NGC 4314, NGC 1433, IC 5240, IC 5340. It is also possible that inner rings as well as a bar can evolve from interlinked rings.

An inner ring formed from a bar will be oval. A circular ring may be formed when the two spiral arms are close enough to perturb each other's orbits such that stars in the inner arm are drawn toward the outer arm on each side of the spiral (NGC 4725, M81, NGC 6902, NGC 7217, NGC 3124). This process may be encouraged if the inner part of the spiral arm has already been sucked into an inner ring or bar. Some galaxies display both types of ring (NGC 1543, IC 1438, NGC 7187). If the ring has sufficient density, stars following the line of the arm and meeting the ring will become gravitationally bound to the ring. The gravitational attraction of the ring will allow it to contain stars with a range of orbital velocities, which will tend to smooth the mass distribution in the ring such that a stable structure may be formed. One would then expect a gap to develop between the ring and the arms (NGC 210, NGC 5701, NGC 1291), or the arms to disappear entirely, where stars from the arms have joined the ring (NGC 7742, PGC 54599, NGC 4553, NGC 4419, NGC 1543, NGC 7020). In the case of PGC 54599, Hoag's Object, one can see a residual arm. The natural evolution of Hoag's Object from a spiral galaxy would explain its symmetrical form and its lack of features characterizing rings resulting from galactic collision. NGC 4725 is sometimes described as a one-armed spiral, but in fact appears to be a ring forming from a two-armed spiral at an earlier stage of evolution than Hoag's object. One arm has almost been destroyed, while the other is still clearly apparent.

## 12. Evolution of Bisymmetric Spirals

Starting from an unstructured initial condition, there is reason to think that mutual gravitational attraction will cause stars to form groups, spiral segments (flocculence), and ultimately to form spiral arms, but no strong reason to expect that these structures will reduce to a bisymmetric spiral. Gas will also be attracted into spiral segments, but when gas clouds meet they combine to form larger clouds, and add more quickly to the mass of the segment, so that the process of reducing spiral segments to a small number of arms is faster for gaseous galaxies.

In the Milky Way, the density of gas in a spiral arm is greater by a factor of about two than that of gas between the arms (figure 14). This is sufficient to maintain spiral structure when gas in the arms meets with outgoing gas. In a symmetric three-armed spiral the density of gas in the arm would be reduced by about 33%, and the density of outgoing gas distributed evenly in the disc would be increased by about 67%. Thus outgoing gas meeting the arm would outweigh the ingoing gas, and would tend to remove gas from the arm. Thus a two-armed gaseous spiral can be stable, whereas multiarmed gaseous spirals cannot.

Outgoing gas applies a pressure to the inside of a spiral arm with an inverse proportionality to radius (figure 15). If one gaseous arm advances compared to the bisymmetric position, the pressure due to gas from the other arm will be reduced. At the same time, pressure on the retarded arm due to outgoing gas from the advanced arm will be increased. Thus gas motions provide a mechanism to maintain the symmetry of two-armed spirals. The same pressure means that gaseous arms will advance with respect to non-gaseous arms and spiral segments. Eventually the arms will combine, so that a two-armed spiral is formed.

Thus gas establishes the pattern of the major spiral arms in flocculent and multi-arm spirals representing earlier stages in the development of grand design spirals. The lower the gas content, the longer the process of evolution. This pattern of evolution is consistent with the observations of Thornley (1996), who found a low-lying spiral structure in four nearby flocculent spirals using near-infrared imaging. We also note that the marked lack of grand design spirals in the Hubble ultra-deep field, in which all galaxies are young, is indicative that this form takes some time to evolve.

The stability of orbits with high eccentricities in the Hercules and Alpha Ceti streams depends on the prior formation of a bisymmetric spiral (figure 17). Francis & Anderson (2009) found that a sharp change in velocity components at age $9\pm1$ Gyrs (previously seen by Quillen & Garnett, 2000) is caused by an increased membership of these streams for stars of greater age. We infer that a grand design two-armed spiral was formed in the Milky Way about 9 Gyrs ago.

### 13. Conclusions

After some time studying the velocity distributions for local stars we have concluded that the observed stellar streams reflect the spiral structure of the Milky Way. We have presented a straightforward model of equiangular spiral arms constructed from elliptical orbits aligned at a focus. This model applies in coordinates rotating at the spiral pattern speed, which is equal to the mean rate of orbital precession. We have shown by qualitative argument and by numerical simulation describing perturbations to elliptical orbits, that, for a range of arm densities, spiral structure is dynamically stable, up to destruction by a bar and/or a ring. We have shown that, for a two-armed equiangular spiral with pitch angle set to match the distribution of neutral hydrogen, the observed eccentricity and velocity distributions are a good fit to the predictions of the model after taking expected perturbations into account. We have accounted for all stellar streams in the observed local velocity distributions. We find that the Sun follows a very typical orbit aligned to the Orion arm, which is a major spiral arm containing Perseus and Sagittarius sectors. We have calculated that its current eccentricity is 0.138. This is a little higher than the modal value, 0.11, for stars in the arm, giving a typical orbital period of about 300 Myrs – longer than usually estimated because of the greater eccentricity. We have seen how spiral structure can evolve to form the rings and bars found in many galaxies, and that gas motions determine that flocculent galaxies evolve toward bisymmetric spirals. We have found that the Milky Way evolved into this form about 9 Gyrs ago.

It is perhaps worth remarking that the model has made genuine predictions, and not merely been retrodictively fitted to data. Having made a prediction of a galactic structure, we searched images to find examples of the configuration. The interlinked ring structure of figure 18 was recognised by the astronomer (E.A.) among the authors, but it was not known to the mathematician (C.F.), who produced the figure from the numerical solution of perturbed orbits. The same was true of the prediction that young stars are to be found on the outside of spiral arms. Nor did we know of galaxies where the spiral arms are separate from the ring. We have not made any predictions of galactic structures for which we were unable to find examples.

### Data

The compiled data used in this paper can be downloaded from
http://data.rqgravity.net/lsr/

17                                *C. Francis, E. Anderson***Benjamin, R**. A. 2008, The Spiral Structure of the Galaxy: Something Old, Something New.... in *Massive Star Formation: Observations Confront Theory,* ASP Conference Series, **387**, ed. Beuther H., Linz H., Henning T., San Francisco, Astronomical Society of the Pacific, p.375-380.
**Bica E., Bonatto C., Barbuy B., Ortolani S.**, 2006, Globular cluster system and Milky Way properties revisited, *A&A*, **450**, 1, pp.105-115.
**Binney J., & Tremaine S.**, 1987, *Galactic Dynamics*, Princeton University Press.
**Chakrabarty D.**, 2007, Phase space structure in the solar neighbourhood, *A&A*, **467**, 145-162.
**Chereul E., Crézé M., Bienaymé O.**, 1998, The distribution of nearby stars in phase space mapped by Hipparcos. II, *A&A*, **340**, 384-396.
**Chereul E., Crézé M., Bienaymé O.**, 1999, The distribution of nearby stars in phase space mapped by Hipparcos. III, *A&A Suppl. Ser.* **135**, 5-8.
**Dehnen W.**, 1998, The Distribution of Nearby Stars in Velocity Space Inferred from Hipparcos Data, *Astron. J.*, **115**, 2384-2396.
**Eggen O.J.**, 1958, Stellar groups. I. The Hyades and Sirius groups, *MNRAS*, **118**, 65-79.
**Eisenhauer F., Genzel R., Alexander T., Abuter R., Paumard T., Ott T., Gilbert A., Gillessen S., Horrobin M., Trippe S., Bonnet H., Dumas C., Hubin N., Kaufer A., Kissler-Patig M., Monnet G., Ströbele S., Szeifert T., Eckart A., Schödel R., & Zucker S.**, 2005, SINFONI in the Galactic Center: Young Stars and Infrared Flares in the Central Light-Month, *Astrophys. J.*, **628**, 246-259.
**Famaey B., Jorissen A., Luri X., Mayor M., Udry S., Dejonghe H., Turon C.**, 2005, Local Kinematics of K and M Giants from CORAVEL/Hipparcos/Tycho-2 Data, *A.&A.* **430**, 165.
**Francis C. & Anderson E.**, 2009, Calculation of the Local Standard of Rest from 20,574 Local Stars in the New Hipparcos Reduction with known Radial Velocities, *New Astronomy*, **14**, pp. 615-629.
**Georgelin Y.M. & Georgelin Y.P.**, 1976, The spiral structure of our Galaxy determined from H II regions, *A.& A.*, 49, 1, p.57-79.
**Goldstein, H.**, 1980, *Classical Mechanics* (2nd ed.). Addison Wesley. pp. 102–105, 421–422.
**Gontcharov G.A.**, 2006, Pulkovo compilation of radial velocities for 35495 stars in a common system, *Astronomy Letters,* **32**, 11, pp. 759–771.
**Hou L.G., Han J.L., Shi W.B.**, 2009, The spiral structure of our Milky Way Galaxy, *A.&A.*, **499**, 473-482.
**Jenkins A.**, 1992, Heating of galactic discs with realistic vertical potentials, *MNRAS*, **257**, 620-632.
**Kalnajs A.J.**, 1973, Spiral structure viewed as a density wave, *Proc. Astr. Soc. Aust.*, **2**, p174 -177.
**Layden A. C., Hanson R. B., Hawley S. L., Klemola A. R., & Hanley C. J.**, 1996, The Absolute Magnitude and Kinematics of RR Lyrae Stars Via Statistical Parallax. *A.J.*, **112**, 2110-2131.
**Levine E.S., Blitz L., Heiles C.**, 2006, The Spiral Structure of the Outer Milky Way in Hydrogen, *Science*, **312**, 5781, pp. 1773-1777.
**C.C. Lin, Yuan, C., and F.H. Shu**, 1969, On the Spiral Structure of Disk Galaxies III. Comparison with Observations, *Ap.J.* **155**, 721-746.
**Mueller, W. K., & Arnett, W. D.**, 1976, Propagating star formation and irregular structure in spiral galaxies, *Ap.J.*, **210**, 670-678.
**Nishiyama S., Nagata T., Sato S., Kato D., Nagayama T., Kusakabe N., Matsunaga N., Naoi T., Sugitani K., and Tamura M.**, 2006, The Distance to the Galactic Center Derived from Infrared Photometry of Bulge Red Clump Stars, *Ap.J.* **647**, pt 1 pp. 1093–1098.
**Oort J.H., Kerr F.J., Westerhout G.**, 1958, The galactic system as a spiral nebula, *MNRAS*, **118**, 379-389.
**Quillen A.C. & Garnett D.R.**, 2000, The saturation of disk heating in the solar neighborhood and evidence for a merger 9 Gyrs ago, arxiv:astro-ph/0004210.
**Reid M.J.**, 1993, The distance to the center of the Galaxy, *ARA&A*, **31**, 345-372.
**Reid M.J. & Brunthaller A.**, 2004, The Proper Motion of Sagittarius A*. II. The Mass of Sagittarius A*, *Ap.J.*, **616**, 872-884.
**Russeil D.**, 2003, Star-forming complexes and the spiral structure of our Galaxy, *A&A*, **397**, 133-146.
**de Simone R.S., Wu X., & Tremaine S.**, 2004, The stellar velocity distribution in the solar neighbourhood, *MNRAS*, **350**, 627-643.
**Sridhar S. & Touma J.**, 1996, Adiabatic evolution and capture into resonance: vertical heating of a growing stellar disc, *MNRAS*, **279**, 1263-1273.
**Thornley M. D.**, 1996, Uncovering Spiral Structure in Flocculent Galaxies, *Ap.J.*, **469**, L45–L48.

## Appendix A  Lindblad's Epicycle Theory

Lindblad's epicycle theory perturbs a circular orbit by superimposing an elliptical motion.

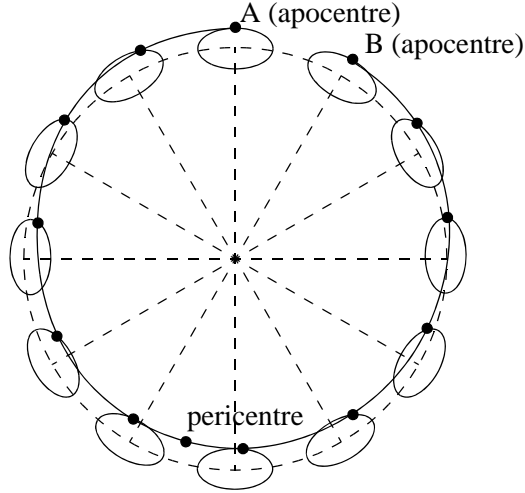

Frequency of circular motion: $\Omega$.
Angular speed, or angular frequency, of circular motion: $2\pi\Omega$.
Frequency of radial oscillation (i.e. elliptical motion): $K$.
Period of radial oscillation: $1/K$.
Angular distance, AB, after $m > 0$ radial oscillations:

$$\mathrm{AB} \;=\; 2\pi n \pm 2\pi\Omega\frac{m}{K},\; n \in \mathbb{Z}.$$

**Note:** Standard treatments (e.g. Binney & Tremaine, 1987; Binney & Merrifield 1998; Carroll & Ostlie, 1996) overlook the possibility of a minus sign.
Frequency of rotating reference frame: $f$.
Angular speed of rotating reference frame: $2\pi f$.
To establish a closed orbit in the rotating reference frame, after $m$ periods of the radial oscillation, we require,

$$2\pi f\frac{m}{K} \;=\; \mathrm{AB} \;=\; 2\pi n \pm 2\pi\Omega\frac{m}{K},$$

$$f \;=\; \frac{n}{m}K \pm \Omega.$$

The usual solution, leading to figure 1, uses plus and $m = -2n$.

$$f \;=\; \Omega - \frac{1}{2}K.$$

This ignores imaging issues with $m > 1$. Spiral pattern speed, as shown in this paper, requires minus, and $m = n = 1$.

$$f \;=\; K - \Omega.$$

For regression of the apocentre, $K < M$ and $f < 0$.

**Appendix B  The Stellar Population**

We derived a stellar population with kinematically complete data by combining astrometric parameters from the recently released catalogue, *Hipparcos, the New Reduction of the Raw Data* (van Leeuwen, 2007a; hereafter "HNR") plus the Tycho-2 catalogue (ESA, 1997) with radial velocity measurements contained in the *Second Catalogue of Radial Velocities with Astrometric Data* (Kharchenko, et al., 2007; hereafter "CRVAD-2"). HNR claims improved accuracy by a factor of up to 4 over the original Hipparcos catalogue (ESA, 1997) for nearly all stars brighter than magnitude 8. Van Leeuwen (2007b) "*confirms an improvement by a factor 2.2 in the total weight compared to the catalogue published in 1997, and provides much improved data for a wide range of studies on stellar luminosities and local galactic kinematics.*" The improvement is primarily due to the increase of available computer power since the original calculations from the raw data, and to an improved understanding of the Hipparcos methodology, which compared positions of individual stars to the global distribution and incorrectly weighted stars in high-density star fields, leading to the well-known 10% error in distance to the Pleiades.

CRVAD-2 contains most of the stars in two important radial velocity surveys: *The Geneva-Copenhagen survey of the Solar neighbourhood* (Nordström, et al., 2004; hereafter G-CS), which surveyed nearby F and G dwarfs, and *Local Kinematics of K and M Giants from CORAVEL* (Famaey et al., 2005; hereafter: "Famaey"). We included about 300 stars in G-CS and Famaey not given in CRVAD-2 and incorporated the revised ages for G-CS II (Holmberg, Nordström and Andersen, 2007). G-CS and Famaey are deemed to be free from kinematic selection bias. The remaining radial velocities in CRVAD-2 are derived from the *General Catalog of Mean Radial Velocities* (Barbier-Brossat and Figon, 2000; hereafter "GCRV") and the *Pulkovo Catalog of Radial Velocities* (Gontcharov, et al. 2006). These are compilations from sources which may contain a selection bias favoring high proper-motion stars (Binney et al., 1997). A number of kinematic studies have concentrated on stars in open clusters. These are likely to be over-represented in CRVAD-2. Since this paper is concerned only with the analysis of bulk motions, selection bias will not have a bearing on our conclusions.

We obtained a population of 20 574 stars by applying the following criteria:

(i) Heliocentric distance within 300 pc based on NRH parallaxes and parallax error less than 20% of parallax. (details below).

(ii) Radial velocity given in CRVAD-2, G-CS or Famaey and uniquely identified to a Hipparcos catalogue number. CRVAD-2 figures were used by default, as CRVAD-2 gives a weighted mean for stars in Famaey having radial velocities from other sources. We excluded stars for which no radial velocity error was given, or for which the quoted error was greater than $5\,\mathrm{km\,s^{-1}}$.

(iii) The object is either a single star or a spectroscopic binary with a computed mean radial velocity. This criterion is determined from flags provided by G-CS, Famaey, Tycho-2, and CRVAD-2.

## Appendix C  Spiral Galaxy Simulation

**http://rqgravity.net/images/spiralmotions/gss.avi.**

   In this animation using 4500 stars, each star follows a rosette. This is the form of orbits predicted under Newtonian gravity for mass distributed symmetrically in the galactic plane and in the halo. Rosettes are aligned by mutual gravity between stars. The gravity of the arm causes stars to follow the arm during the ingoing part of their orbit. The simulation uses orbits with random eccentricities between 0.10 and 0.18, corresponding to observations of local stars in the Milky Way. The pattern created is a grand-design two-armed spiral. To see an orbit, follow the path of one of the giant stars. The spiral pattern seems to shrink, but really it is rotating slowly retrograde to stellar orbits. If one were to scale this galaxy to the Milky Way the Solar orbit would be about midway in the spiral. The diameter of the galaxy would be about 130,000 light years. The Sun has been moving down the Orion arm for about 150 million years, and is now crossing outwards through this arm, prior to leaving the arm, crossing the Centaurus arm, and rejoining the Orion arm. The relationships between speed of rotation of the bar (bar pattern speed) the speed of rotation of the spiral (spiral pattern speed) and orbital velocities depend on the mass distribution of the Galaxy and are not known.

## Appendix D Milky Way Map

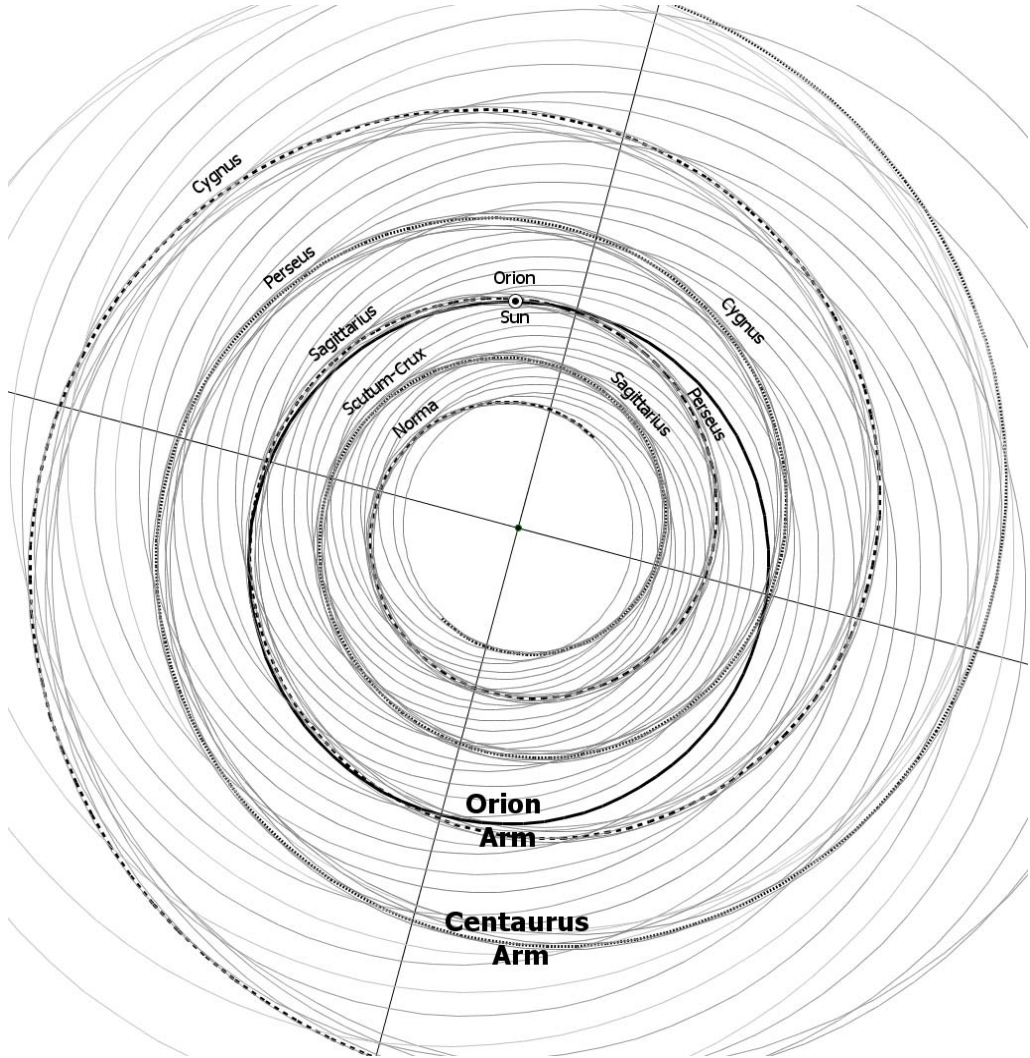

**Figure 19:** Two-armed spiral model of the Milky Way, based an angular increment of $\tau = 30°$ for each 105% enlargement, giving a pitch angle of 5.44°. The solar eccentricity, 0.138, has been used for the diagram. Good fits are obtained with eccentricities from about 0.1 to about 0.2. A two-armed spiral necessitates a little care to avoid confusion in naming the arms, because traditionally named sectors with the same name lie on different arms. Orion is not a separate spur, but is a part of a major arm connecting Perseus in the direction of rotation to Sagittarius in the direction of anti-rotation. We have called the Orion Arm the major spiral arm containing the Sun as well as Norma, Perseus, Orion, Sagittarius, and Cygnus sectors. The Centaurus arm contains Sagittarius, Scutum-Crux, Cygnus, and Perseus sectors. The solar orbit is shown in approximation, together with the major axis and latus rectum.

### Appendix E  Perturbation to the Eccentricity Vector

If we regard the mass distribution of the Galaxy as equivalent to a central mass plus spiral structure, then the spiral structure will perturb elliptical orbits. The eccentricity vector is defined as

$$\boldsymbol{e} = \frac{|v|^2 \boldsymbol{r}}{\mu} - \frac{(\boldsymbol{r} \cdot \boldsymbol{v})\boldsymbol{v}}{\mu} - \frac{\boldsymbol{r}}{|\boldsymbol{r}|}.$$

For a Keplerian orbit, $\boldsymbol{e}$ is a constant of the motion. In coordinates rotating at the rate of precession, the orbit is an oval and the variation in $\boldsymbol{e}$ is cyclic. We may ignore small differences between an oval and an elliptical orbit since we are concerned with the locus of the orbit, not obedience to Keplerian motion. The difference between an oval orbit in rotating coordinates and an ellipse is not sufficient to alter the spiral pattern seen in figure 12. We will thus assume that $\boldsymbol{e}$ is approximately constant in rotating coordinates.

Using primes to denote the perturbation due to the arms,

$$\boldsymbol{e} + \boldsymbol{e}' = \frac{|v + v'|^2(\boldsymbol{r} + \boldsymbol{r}')}{\mu} - \frac{((\boldsymbol{r} + \boldsymbol{r}') \cdot (\boldsymbol{v} + \boldsymbol{v}'))(\boldsymbol{v} + \boldsymbol{v}')}{\mu} - \frac{\boldsymbol{r} + \boldsymbol{r}'}{|\boldsymbol{r} + \boldsymbol{r}'|}.$$

Neglecting squares in the perturbation to $\boldsymbol{v}$ and the perturbation to $\boldsymbol{r}$ (since the motion is approximately elliptical),

$$\boldsymbol{e}' \approx \frac{2(\boldsymbol{v} \cdot \boldsymbol{v}')\boldsymbol{r}}{\mu} - \frac{(\boldsymbol{r} \cdot \boldsymbol{v})\boldsymbol{v}'}{\mu} - \frac{(\boldsymbol{r} \cdot \boldsymbol{v}')\boldsymbol{v}}{\mu}.$$

Writing $\boldsymbol{w} = \boldsymbol{v}'$, and using carets to denote unit vectors and unbold font to denote magnitude, $\boldsymbol{r} = r\hat{\boldsymbol{r}}$, $\boldsymbol{v} = v\hat{\boldsymbol{v}}$, $\boldsymbol{w} = \boldsymbol{v}' = w\hat{\boldsymbol{w}}$,

$$\boldsymbol{e}' \approx \frac{wrv}{\mu}(2(\hat{\boldsymbol{v}} \cdot \hat{\boldsymbol{w}})\hat{\boldsymbol{r}} - (\hat{\boldsymbol{r}} \cdot \hat{\boldsymbol{v}})\hat{\boldsymbol{w}} - (\hat{\boldsymbol{r}} \cdot \hat{\boldsymbol{w}})\hat{\boldsymbol{v}}).$$

For orbits of low eccentricity, the variation in $rv$ is low. So, after carrying out the products and sums of the unit vectors, the magnitude of the perturbation is determined principally by $w$, and is given by the integral over time of the acceleration due to the spiral arm. In a typical orbit there are six phases of perturbation due to spiral structure, shown in figure 20, numbered starting from the point at which a star joins the arm, shortly before apocentre. The qualitative effect of the various perturbations in a typical orbit is seen in figure 21.

When a star approaches pericentre, it will typically be near the inside of the arm. Such stars are close to the Sun, and appear prominently on the velocity distribution, but typically have lower eccentricities than the Sun. The star then leaves the arm, crossing the second arm close to the semi-latus rectum. The dominance of the Hyades stream in the velocity plots is due to the fact that all stars with orbits aligned to the Centaurus arm cross the Orion arm at a similar point in their orbits.

A star approaches apocentre relatively slowly, at which point it is also approaching the arm to which its motion is tied. Because the star spends longer near apocentre, the perturbation during this part of the orbit produces larger shifts in both magnitude and direction of the eccentricity vector. The eccentricity vector advances at this point of the orbit, and stars pass through the Solar neighbourhood, leading to another peak in the observed distribution. Once a star passes through the arm it is too far away to appear in the local velocity distribution, so the distribution here is sparse.

**Figure 20:** Perturbations to a typical orbit due to spiral structure.

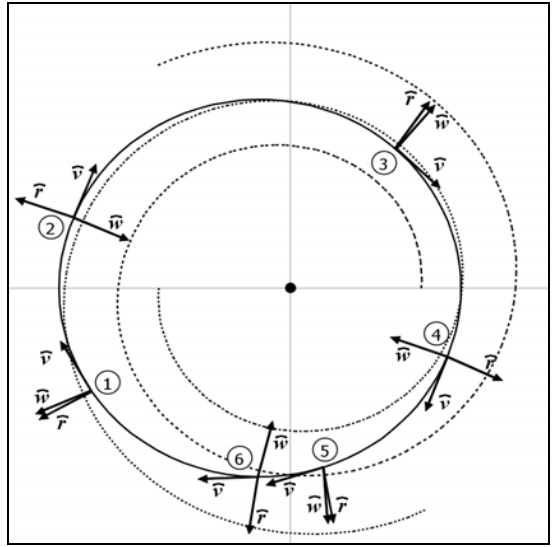

**Figure 21:** Result of perturbations due to spiral structure on a typical orbit. The unit vectors from figure 20 are shown in the first column, and their products evaluated in the next three columns. In the fourth column they are summed and multiplied by a factor which depends on $w$, to give the perturbation to eccentricity due to each phase of the motion. In the final column they are added to the eccentricity vector. The actual effect of the perturbation depends critically on the initial eccentricity vector and orbital radius. Within bounds, if $e$ is low, then the perturbation in phase 1 is high, and $e$ will increase over an orbit. If $e$ is high the duration of phase 1 will be short, and the orbit will not go deep into the arm. There will then be a low perturbation from phases 1 and 2, and a net decrease in $e$ over an orbit.

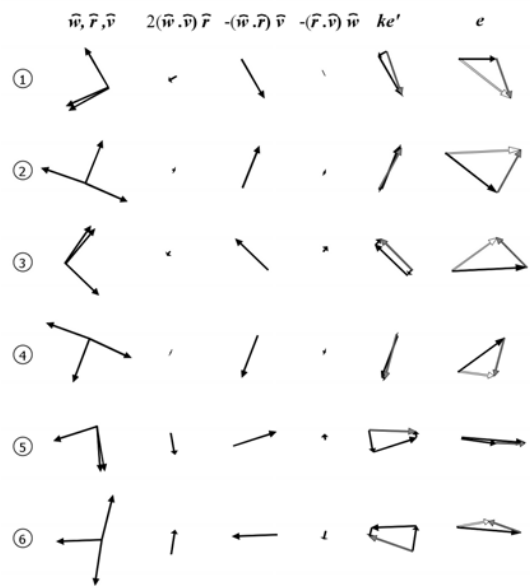

**Figure 22:** The velocity distribution for giants with $B - V > 0.4$ mag and G-CS dwarfs with ages 2-13 Gyrs. Young stars in the Pleiades stream are effectively removed, but there is a distribution of stars with slightly higher eccentricities near apocentre joining the arm. The other streams are all seen in the plot. The dearth of stars after apocentre is due to the fact that these stars tend to be on the outside of the arm. The Hyades cluster is thought to be about 600 Myrs old and is over-represented in the plot. It is possible that the alignment of orbits in the arm over a substantial time period has enabled the Hyades cluster to capture a number of old stars.

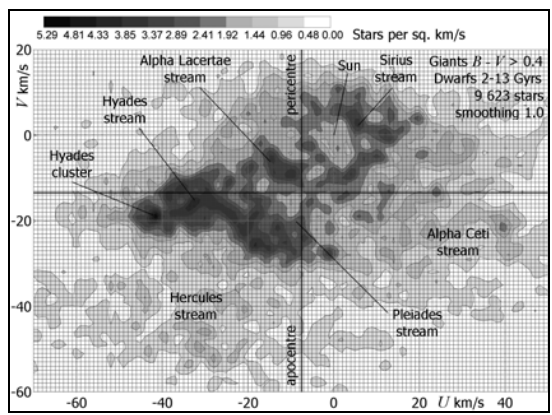

**Figure 23:** The eccentricity distribution for giants with $B-V > 0.4$ mag and G-CS dwarfs with ages 2-13 Gyrs. Qualitative perturbation to the eccentricity vector is shown for a typical orbit. During phase 1, stars rejoining the arm, eccentricity increases, and the eccentricity vector advances, so there is a high density on this part of the plot. During phase 2 stars lie on the outside of the arm, and the eccentricity vector regresses, so there is a low density on the plot. Density increases at phase 3, at which point stars lie on the inside of the arm near the Sun. The high density for the Hyades stream results from the fact that stars with orbits aligned to the Centaurus arm cross the Orion arm at a similar point in the outward part of their orbit.

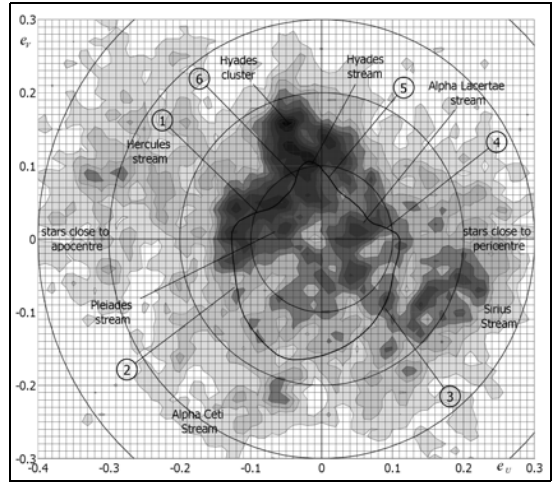

**Figure 24:** Orbits (left) and changes in eccentricity in *U-V* coordinates (right) using a numerical model based on a central mass with perturbations due to spiral structure, intended as an approximation to motion due to a distributed mass in coordinates rotating at the rate of precession of apocentre. In each plot the star starts at apocentre with initial orbital parameters to reflect spiral arm motions. Eccentricity in the first orbit starts low and becomes high. The second orbit is fairly stable. In the third orbit eccentricity starts high and becomes low. The fourth orbit starts poorly aligned to the disc, and over corrects before settling back. The first and third orbits also naturally precess to better fit the arm.

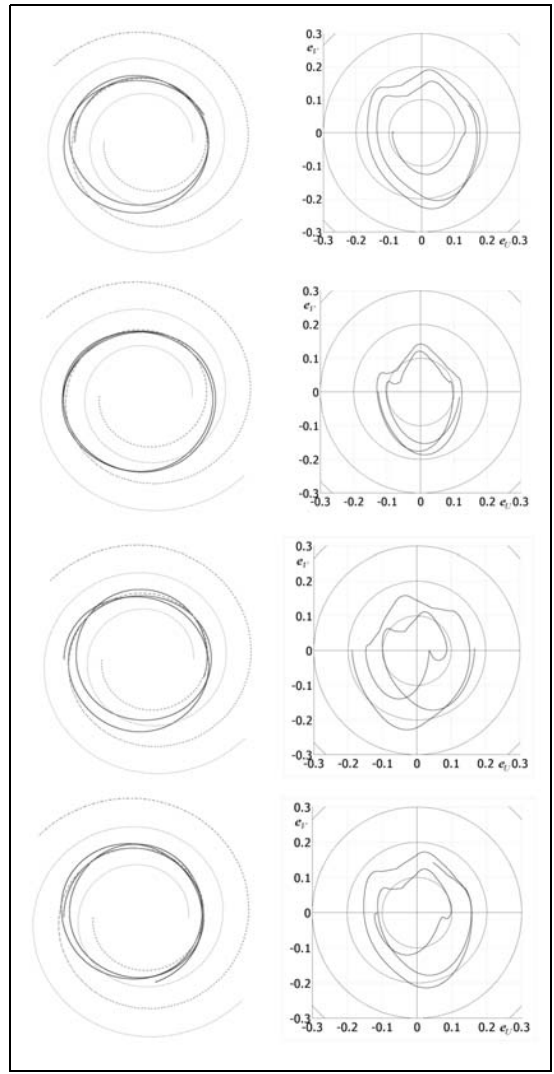

We plotted the velocity distribution for mature orbits by restricting the population to F&G dwarfs with ages 1-13 Gyrs given in G-CS II and giants redder than $B - V = 0.4$ mag (figure 22) and the corresponding eccentricity distribution (figure 23). *U-V* coordinates rotate through an orbit, so that, if unperturbed, the eccentricity vector would trace a circle. We used estimates of the perturbation to fit a qualitative plot of the eccentricity vector in a typical orbit to the eccentricity distribution for mature orbits.

For a symmetric mass distribution the first order effect of the distributed mass in the disc and in the halo is to the precession of apocentre. This effect is removed in coordinates rotating with spiral pattern speed. The resulting oval is approximated by an ellipse. We modelled perturbed elliptical orbits using a numerical solution to the Newtonian equations for a central mass plus spiral structure, intended as an approximation to the correct equations in coordinates rotating with spiral pattern speed. A more accurate model would include further perturbations and would explicitly calculate spiral pattern speed, but mass models given in literature (e.g. Klypin, Zhao, and Somerville, 2002) do not reveal that the mass distribution of the Milky way is sufficiently well known for a precise solution for the Milky Way spiral. Our numerical solution is intended only to demonstrate that perturbations due to spiral structure cause adjustments to eccentricity and orientation of apocentre such that alignment with the arm tends to improve over time. It uses parameters approximate to those of the Milky Way showing a qualitative fit between the predictions of the model and the observed distribution, and illustrates the general principle that, for a range of real matter distributions, perturbations due to spiral structure have the effect of stabilizing the spiral.

A central mass of 77 billion solar masses was chosen to match the speed of the LSR at an adopted radius of 7.4 kpc. This is a little higher than the true enclosed mass on account of the distribution of matter in the disc. For the perturbation we used an inverse acceleration law,

$$a = -\frac{\mu_{\text{arm}}}{d} \text{ for } d > 0.04R,$$

and a linear law within a nominal arm width,

$$a = -\frac{\mu_{\text{arm}}d}{(0.02R)^2} \text{ for } d \leq 0.04R,$$

where $R$ is distance to Sgr A* and $d$ is the distance to the centre of the nearest spiral arm. A step size $6 \times 10^{11}$ secs (19 013 years) was chosen such that an unperturbed orbit did not expand noticeably. $\mu_{\text{arm}}$ was taken to be constant, since we are interested in the perturbation at the Solar radius. Values in the range $1 \times 10^8 < \mu_{\text{arm}} < 2 \times 10^8$ m$^2$s$^{-2}$ gave reasonable qualitative fits with the observed distribution. $\mu_{\text{arm}} = 2 \times 10^8$ m$^2$s$^{-2}$ was used in the plots. This is perhaps on the high side of the true value. We ran the model for a number of initial conditions, starting with stars at apocentre, and plotted the orbit and the eccentricity vector. The model confirmed that, for a broad range of arm densities, orbits do not repeat identically, but vary between lower and higher eccentricities and precess to better match the locus of the arm (figure 24).